\documentclass[conference]{IEEEtran}
\IEEEoverridecommandlockouts
\usepackage{cite}
\usepackage{amsmath,amssymb,amsfonts}
\usepackage{algorithmic}
\usepackage{graphicx}
\usepackage{textcomp}
\usepackage{xcolor}

\begin{document}

\title{Spatiotemporal Prediction of Electric Vehicle Charging Load Based on Large Language Models
}

\author{\IEEEauthorblockN{Hang Fan$^{1}$, Mingxuan Li$^{2}$, Jingshi Cui*$^{3}$, Zuhan Zhang$^{1}$,  Wencai Run$^{1}$ and Dunnan Liu$^{1}$}
\IEEEauthorblockA{$^{1}$School of Economics and Management, North China Electric Power University, Beijing, China\\
$^{2}$Department of Electrical Engineering, Tsinghua University, Beijing, China\\
$^{3}$Department of Control Science and Intelligence Engineering, Nanjing University, Nanjing, China \\
*jscui@nju.edu.cn}
}

\maketitle
\begin{abstract}
The rapid growth of EVs and the subsequent increase in charging demand pose significant challenges for load grid scheduling and the operation of EV charging stations. Effectively harnessing the spatiotemporal correlations among EV charging stations to improve forecasting accuracy is complex. To tackle these challenges, we propose EV-LLM for EV charging loads based on LLMs in this paper. EV-LLM integrates the strengths of Graph Convolutional Networks (GCNs) in spatiotemporal feature extraction with the generalization capabilities of fine-tuned generative LLMs. Also, EV-LLM enables effective data mining and feature extraction across multimodal and multidimensional datasets, incorporating historical charging data, weather information, and relevant textual descriptions to enhance forecasting accuracy for multiple charging stations. We validate the effectiveness of EV-LLM by using charging data from 10 stations in California, demonstrating its superiority over the other traditional deep learning methods and potential to optimize load grid scheduling and support vehicle-to-grid interactions.
\end{abstract}


\section{Introduction}
The transition to a low-carbon economy has underscored the importance of EVs as a promising solution to reduce carbon emissions and mitigate fossil energy shortages \cite{starke2023improving}. In 2024, the global adoption of EVs has increased significantly, with sales surpassing 10 million units between January and August, which accounts for $18\%$ of total new car sales, and shows an increase of approximately $3\%$ compared to 2023. 

However, this rapid growth in EV adoption, coupled with the escalating demand for charging, poses substantial challenges for power grid management. The variability in EV charging loads can disrupt the stable operation of distribution networks, complicating efforts for optimization and scheduling. Moreover, EV manufacturers and charging infrastructure providers are increasingly participating in vehicle-to-grid interactions and electricity markets as load aggregators, as exemplified by initiatives such as Tesla's Auto-Bidder. Consequently, accurate prediction of EV charging loads is essential to effectively address these challenges.

Research on EV charging load forecasting is divided into model-driven and data-driven methods. Model-driven methods involve establishing mathematical models to deduce user behavior under various factors and simulate the spatiotemporal distribution characteristics of charging loads. \cite{su2017ev} proposed a dynamical evolution model for the spatial and temporal distribution of EV’s charging demands based on Agent-cellular automata. \cite{multi_info} proposed an EV charging load forecasting method that considers the real-time interaction of multi-source information and user regret psychology. The authors used the Monte Carlo method to simulate the travel and charging processes and obtain the spatiotemporal distribution of charging loads within the area.

With the development of big data, the data-driven EV load forecasting methods have gradually gained attention. \cite{yaofang} proposed a novel ISSA-CNN-GRU model to combine the feature extraction capabilities of CNN with the time series forecasting capabilities of GRU and optimized hyperparameters using an improved sparrow search algorithm for EV charging load forecasting. In response to the uncertainty and long-term forecasting issues, \cite{yao2024ev} proposed a new algorithm based on dynamic adaptive graph neural networks. The algorithm effectively captures the spatiotemporal fluctuations of charging loads through spatiotemporal correlation feature extraction which enhances the model's forecasting ability. Experimental results show that the method achieved more accurate load forecasting on the Palo Alto dataset. \cite{wang2023predicting} introduced a heterogeneous spatial-temporal GCN designed to predict EV charging demand at various spatial and temporal resolutions, which achieved this by constructing heterogeneous graphs that capture the spatial correlations between different charging regions. With the rise of generative artificial intelligence technologies, the application of large language models for time series forecasting has begun to attract attention\cite{jin2023time}. For example, \cite{wu2024stellm} utilized a pre-trained LLM for wind speed prediction and proved its superiority over other traditional deep learning models which indicates its applicability for energy time series.


Although existing works have proposed various EV load forecasting methods, we find that existing methods often struggle to adapt to the dynamic spatial-temporal relationship of user charging behavior, which is influenced by multiple factors, including time, location, and external conditions. To address these limitations, this paper proposes a novel approach combining LLMs and GCNs to enhance spatiotemporal prediction tasks for EV charging loads. By leveraging the capabilities of LLMs, we aim to better model the intricate dynamics of charging behavior compared to traditional methods. Our approach seeks to improve the understanding and forecasting of evolving charging demand patterns, ultimately providing a robust solution for load grid management and facilitating market participation by load aggregators.

In summary, the contributions of our proposed EV load prediction method can be summarized as follows:
\begin{itemize}
\item Adoption of LLMs: To our best knowledge, this paper is the first one to combine LLMs and GCNs to enhance EV charging load forecasting, presenting a novel approach that diverges from traditional and solely data-driven methodologies.
\item Enhanced Prediction Accuracy: By integrating GCNs with LLMs, our method effectively captures complex spatiotemporal dependencies and multi-variable interactions, addressing the limitations of existing models that struggle with long sequence dependencies.
\item Reprogramming and Embedding Innovations: We introduce a unique reprogramming layer designed to efficiently map time series data into a format suitable for LLMs processing, thereby enhancing the model's ability to interpret and predict EV charging patterns.
\end{itemize}

\section{Problem Description and Model Structure}
\subsection{Problem Description}
Forecasting the load at EV charging stations is fundamentally a time series forecasting problem that can be modeled uniformly. The charging load in the next $H$ time steps can be predicted using historical charging load data from the previous $M$ time steps and numerical weather prediction (NWP) data for the next $N$ steps. This relationship can be formulated as:
\begin{equation}
\begin{aligned}
&P_{t+1},\ldots P_{t+H}=\\
&f\left(P_{t-1},\ldots P_{t-M},{NWP}_{t+1},\ldots{NWP}_{t+N},\Lambda \right),
\end{aligned}
\end{equation} 
where $P_{t} \in R^{n \times 1}$ represents the vector of EVs chargining load, and $N W P_{t} \in R^{n \times k}$ denotes the NWP variables. Here, $n$ is the number of EV charging stations, $k$ is the number of the NWP variables, $\Lambda$ is the adjacent matrix about the correlation among stations, and $f$ is the prediction model.

\subsection{Overall Architecture of EV-LLM Model}
\begin{figure*}[htbp]
                \centering              \includegraphics[width=0.9\linewidth]{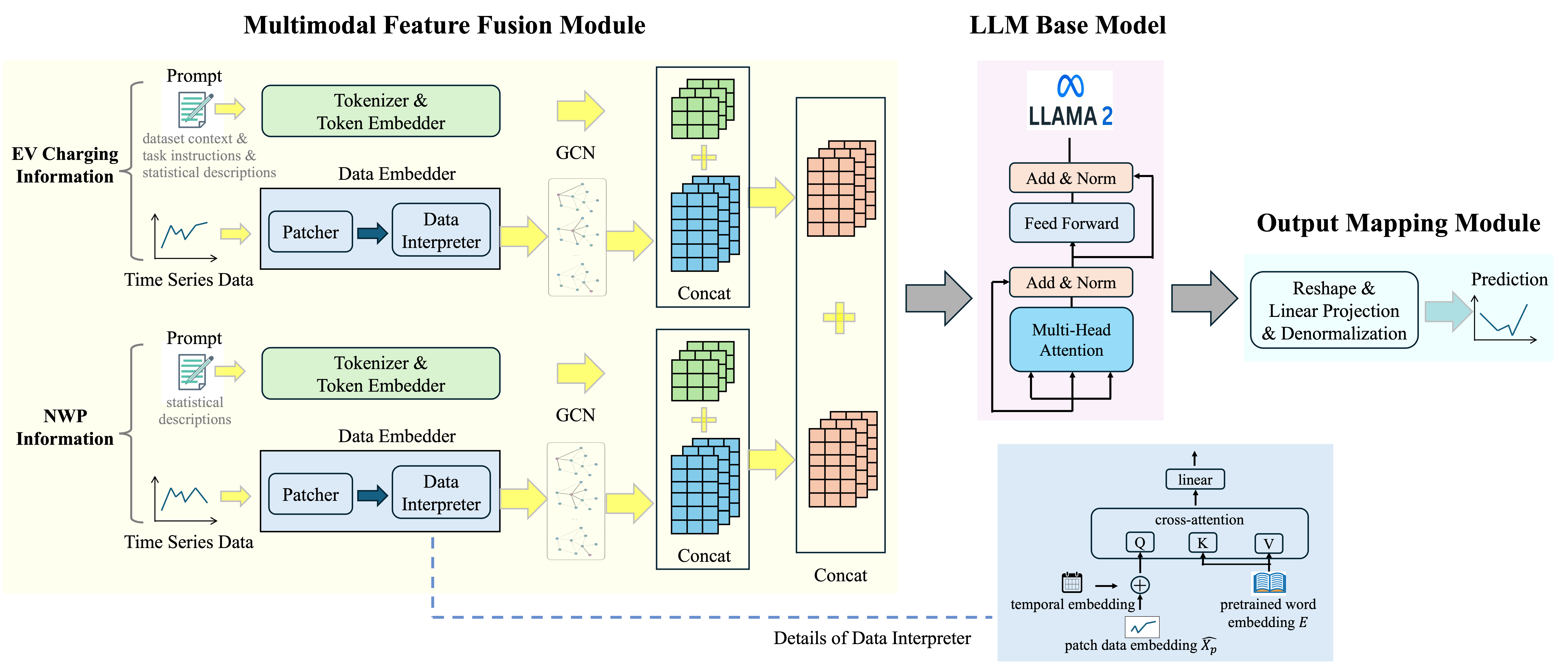}
                \caption{Overall Architecture of EV-LLM Model}
                \label{fig:overall}
        \end{figure*}

In this paper, we introduce the EV-LLM model, a framework for EV load prediction based on LLMs. This approach enhances the LLM's reasoning capabilities for time series by utilizing pretrained word embeddings and prompts to improve task comprehension. By employing customized multimodal inputs and performing minimal parameter training on the LLMs, the proposed model demonstrates exceptional abilities in EV load prediction tasks. The architecture of proposed EV-LLM model is illustrated in Fig. \ref{fig:overall} and consists of three primary components. While LLMs are computationally intensive, to reduce training cost we focused on training specific components to better align with the requirements of spatiotemporal prediction task instead of tuning the entire model.

\subsubsection{Multimodal Feature Fusion Module} This module aligns textual prompts with reconstructed time series data, embedding the time series using projections from pretrained LLMs. By reshaping the data into a format comprehensible to the LLMs, it maps the time series into the same high-dimensional space as the prompt embeddings which enables the LLMs to interpret the prediction task more effectively. Detailed discussions regarding this module can be found in Section \ref{Section:Multimodal}.

\subsubsection{LLMs Base Model}
With extensive pre-training across diverse tasks, LLMs exhibit strong generalization capabilities. Previous studies have leveraged pre-trained LLMs for time series forecasting, suggesting similar methods could enhance accuracy in charging station power forecasting. Given the computational demands of LLM inference and task complexity, this paper selects LLAMA2 as the base model.

\subsubsection{Output Mapping Module} 
In the output layer, the EV-LLM maps the generated output vector into the required format for time series predictions, producing the final forecasting data. This mapping process involves flattening the predicted results from the LLMs, applying a linear projection, and then denormalizing the output into the desired format (e.g., standard prediction data), yielding the final predicted result $Y^\ast$.

For these modules, the multimodal feature fusion module is most critical and will be discussed in detail in Section \ref{Section:Multimodal}.

\section{Multimodal Feature Fusion Module} \label{Section:Multimodal}
The Multimodal Feature Fusion Module aims to integrate diverse input data for compatibility with the LLMs, including three key components: prompt configuration, data embedder, and GCN.

\subsection{Prompt Configuration}
Appropriate prompt prefixes should be designed to enhance the LLM's understanding and performance in predicting time series data. By integrating prompt prefix optimization with reprogrammed time series blocks as the overall input layer, we can guide the model’s focus on specific tasks and data characteristics. In the context of EV charging load prediction, data from multiple stations is decomposed into various representation prompts. These prompts utilize carefully designed words to represent the spatiotemporal correlations between stations. For tasks exhibiting strong spatiotemporal characteristics, we focus on three key aspects for crafting prompt prefixes: dataset context, task instructions, and statistical descriptions.

\subsubsection{Dataset Context} This aspect provides essential background information relevant to EV load prediction, including data sources, collection frequency, geographical locations, and environmental conditions. For instance: ``This dataset consists of historical charging records from an EV station located at 30 degrees north latitude and 120 degrees east longitude. Data is collected hourly, capturing charging load and corresponding weather conditions over the past year, including sunny, cloudy, and rainy days." Providing this context helps the LLM establish a foundational understanding of the data, clarifying its physical significance and practical applications.

\subsubsection{Statistical Descriptions} This component includes insights into data trends, periodic characteristics, fluctuation ranges, and time lags. For example: ``Daily charging load exhibits clear daily periodic fluctuations, with higher usage in the morning and afternoon, and lower levels during midday and evening. Notably, there is a significant weekend fluctuation in charging volume, with increased usage on weekdays and decreased usage on weekends." Summarizing these characteristics aids the LLMs in better comprehending and processing the time series data.

\subsubsection{Task Instructions} These instructions clarify the specific goals and requirements for the LLMs within the current task. For example: ``Please predict the changes in EV charging load over the next 48 hours based on the historical charging data and weather conditions from the past week." Clear task instructions enable LLM to adapt its focus to various downstream tasks, ensuring it is calibrated for specific requirements.

By combining the prompt prefix with the reprogrammed vector and inputting it into the LLMs, we can effectively reduce prediction errors and enhance overall performance.

\subsection{Data Embedder}
The data embedder reconstructs time series data to improve compatibility with the LLMs, consisting of two main components: the data patcher and data interpreter. The data patcher segments volatile EV charging data, while the data interpreter maps it into a high-dimensional feature space.

\subsubsection{Data Patcher} Due to the high volatility in EV charging load data, slicing it into various segments aids in capturing correlation patterns. This process includes two main steps:

\begin{itemize}
    \item Instance Normalization: Each input channel is normalized independently, achieving zero mean and unit variance. This process counteracts distribution shifts across different periods, essential for EV load prediction where data can vary significantly due to factors like weather.
    \item Patch Segmentation: Using a sliding window, the time series \( X^{(i)} \) for feature $i$ is divided into continuous patches of length \( L_p \). These patches (with a total of \( P \)) capture both short-term and long-term variations, crucial for improving prediction accuracy in the presence of daily and seasonal patterns in EV load data \cite{ilharco2022patching}. Concatenating each patch results in \( X_P^{(i)} \in \mathbb{R}^{P \times L_p} \).
\end{itemize}

\subsubsection{Data Interpreter} Data interpreter maps low-dimensional time series slices into high-dimensional embeddings aligning with the prompt prefix embeddings, serving as the input for the LLMs. The data interpreter utilizes the attention mechanism to extract the relationship between time series data and word vectors, effectively ``translating" the time series data into a format suitable for prediction\cite{asudani2023impact}.

Each time series patch finds its corresponding mapping from the synchronized input of the LLM's text set via the multi-head self-attention mechanism. This mechanism captures features of the time series data from various levels and aspects, using attention weights to identify key predictions and encoding this information as natural language. For instance, a time series patch might describe the EV charging load changes as ``a short-term increase followed by a steady decline", which intuitively conveys the dynamic behavior of the time series. 

In the EV charging load prediction task, the reprogramming layer is structured as follows: For each feature \( i \), the patch data \( X_P^{(i)} \) is initially passed through a linear layer, embedding each patch of length \( L_p \) into the hidden dimension \( d_m \) of the attention block, yielding \( \hat{X}_P^{(i)} \in \mathbb{R}^{P \times d_m} \). Next, the query matrix is computed as \( Q_k^{(i)} = \hat{X}_P^{(i)} W_K^Q \), the key matrix as \( K_k^{(i)} = E W_k^K \), and the value matrix as \( V_k^{(i)} = E W_k^K \), where \( E \) represents the pretrained word embeddings of the LLM, provided by the LLM's token embedder. Subsequently, the charging load time series segments are reprogrammed within each attention head as follows:
\begin{equation}
\begin{aligned}
Z_k^{\left(i\right)}&=Attention\left(Q_k^{\left(i\right)},K_k^{\left(i\right)},V_k^{\left(i\right)}\right)\\
&=Softmax\left(\frac{Q_k^{\left(i\right)}K_k^{\left(i\right)T}}{\sqrt{d_k}}\right)V_k^{\left(i\right)}.    
\end{aligned}
\end{equation}

The outputs of each head are combined to form \( Z^{(i)} \in \mathbb{R}^{P \times d_m} \), which is then linearly projected to match the hidden dimension $d_{\text{LLMs}}$ of the LLMs base model, resulting in the reprogrammed output \( O^{(i)} \in \mathbb{R}^{P \times d_{\text{LLMs}}} \). This reprogramming process enables the LLMs to efficiently interpret and handle time series data, adaptively capturing key features of EV load variations and improving prediction accuracy.

\subsection{Graph Neural Network}
Following the data embedding, the charging load and weather characteristic data of each charging station are represented as vectors of time series. GCN is increasingly recognized for its ability to extract spatiotemporal features in EV charging load forecasting \cite{kim2024spatial}. To effectively capture spatiotemporal correlations, we employ two layers of GCN for data transformation.The feature propagation from layer \( l \) to layer \( l+1 \) in the GCN can be represented as follows:
\begin{equation}
\mathbf{X}^{(l+1)} = \sigma \left( \mathbf{A} \mathbf{X}^{(l)} \mathbf{W}^{(l)} \right),
\end{equation}
where $\mathbf{X}^{(l)}$ is the feature matrix at layer $l$, which is also the output of data embedder. Here, $\mathbf{A}$ is the graph convolutional matrix, $\mathbf{W}^{(l)}$ is the weight matrix at layer $l$, and $\sigma$ is a non-linear activation function such as ReLU function.

The key aspect of the GCN is the construction of the graph, for which we utilize the mutual adjacency matrix (MAM) method as described in \cite{kim2024spatial} and further details can be found in \cite{kim2024spatial}. The input to the GCN consists of embedding features for each node, derived from the data embedder. By concatenating the output from the GCN module with the token embedder, we obtain a matrix with dimensions $\mathbf{N\times d}$, where $N$ represents the number of stations, and $d$ denotes the feature dimensions. Next, we concatenate this matrix of EV charging information with the NWP data as part of the multimodal feature fusion module. This concatenated data is then flattened into one-dimensional arrays, which serve as the input for the LLMs.

In this way, the GCN transforms spatiotemporal data into a format that is suitable for processing by LLMs. The synergy between the spatiotemporal features extracted by the GCN and the generalization capabilities of the LLMs enables the model to effectively manage multimodal and multidimensional datasets, significantly enhancing prediction accuracy.




\section{Case Study}
\subsection{The Training Dataset and Experiment Environment}
\subsubsection{Dataset}In this study, we utilize public charging data from Palo Alto, California, to demonstrate the effectiveness of our proposed method. The dataset consists of one year (2020) of EV charging records from 10 charging stations, which we use for training, validation, and testing. Additionally, we collect open-source weather data, including temperature, precipitation, humidity, and wind speed, from online sources to enhance the prediction model. The entire dataset is divided into training (70\%), validation (20\%), and test (10\%) sets.

\subsubsection{Experiment Environment}The pre-trained base model used in this study is LLAMA2-7b, implemented with PyTorch. All experiments were conducted on Ubuntu 18.04 LTS, utilizing an NVIDIA A100 GPU. Through grid search, the patch length $L_p$ is determined to be $16$, resulting in a total of $P = 12$ patches. The hidden dimension $d_m$ in the data embedder is set to $32$. Training concluded after $36$ epochs, using an early stopping mechanism based on the validation set, and take a total of $5,031$ seconds.

\subsection{The Statistic Prediction Results of Different Methods}
To validate the performance of our prediction model, we compare against state-of-the-art methods\cite{yaofang, yao2024ev, kim2024spatial} as well as traditional time series prediction models, including Transformer, GCN, LSTM, Informer, and Autoformer. Given the presence of multiple charging stations and varying prediction time steps, we calculate the average Root Mean Square Error (RMSE) and Mean Absolute Error (MAE) across all stations and time steps\cite{kim2024spatial}. The results are summarized in Table~\ref{tab:results}.

\begin{table}[ht]
  \centering
  \caption{Predicton Performance Analysis of Different Methods}
  \label{tab:results}
  \begin{tabular}{c|c|c} 
    \hline 
    Model & Average MAE& Average RMSE  \\ \hline 
    EV-LLM& 1.7288& 2.8960\\ 
    ISSA-CNN-GRU\cite{yaofang} & 2.1596& 3.1819\\
    AST-GCN\cite{yao2024ev} &1.9711 & 3.2103\\ 
    mRGC-CBi-LSTM\cite{kim2024spatial} & 1.9319& 3.0622\\
    Autoformer& 2.0676 & 3.0398\\ 
    Informer &2.0772 &3.4724\\ 
    Transformer& 2.4010& 3.7109\\
    GCN& 2.5671&3.7715\\
    LSTM& 2.4549&3.7435\\ \hline
  \end{tabular}
\end{table}

\begin{figure}
    \centering
    \includegraphics[width=0.97\linewidth]{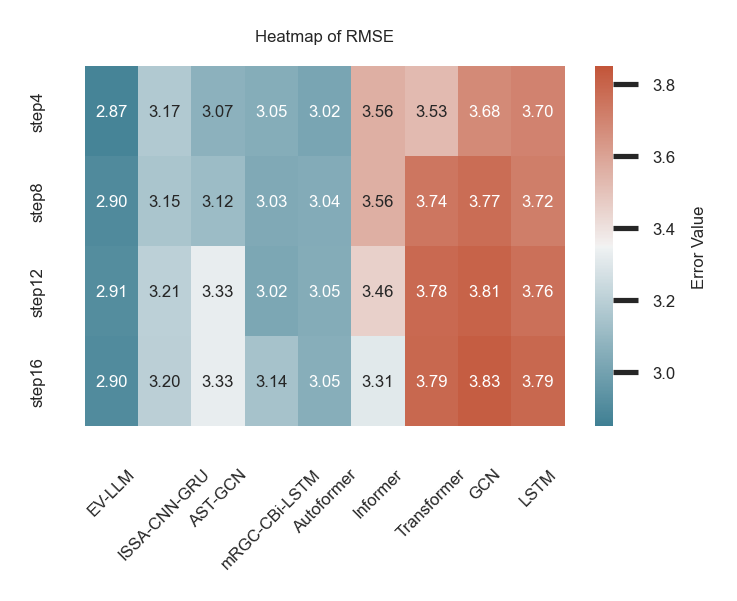}
    \caption{Model prediction RMSE comparison}
    \label{fig:enter-label}
\end{figure}

From the average values across all charging stations and time steps, we observe that the EV-LLM outperforms the other models, exhibiting lower average MAE and RMSE. Notably, methods that consider charging load characteristics, such as those in \cite{yaofang}, \cite{yao2024ev}, and \cite{kim2024spatial}, outperform traditional time series methods like the Transformer. To further illustrate the effectiveness of our approach, we compare the prediction results of benchmark methods across different time steps, presenting heatmap visualizations in Fig.~\ref{fig:enter-label}. The proposed EV-LLM consistently shows lower RMSE and MAE across various time steps, indicating superior performance compared to other deep learning methods.

\subsection{Visualization of Prediction Results on Charging Station}
\begin{figure}[!htbp]
                \centering
                \includegraphics[width=1\linewidth]{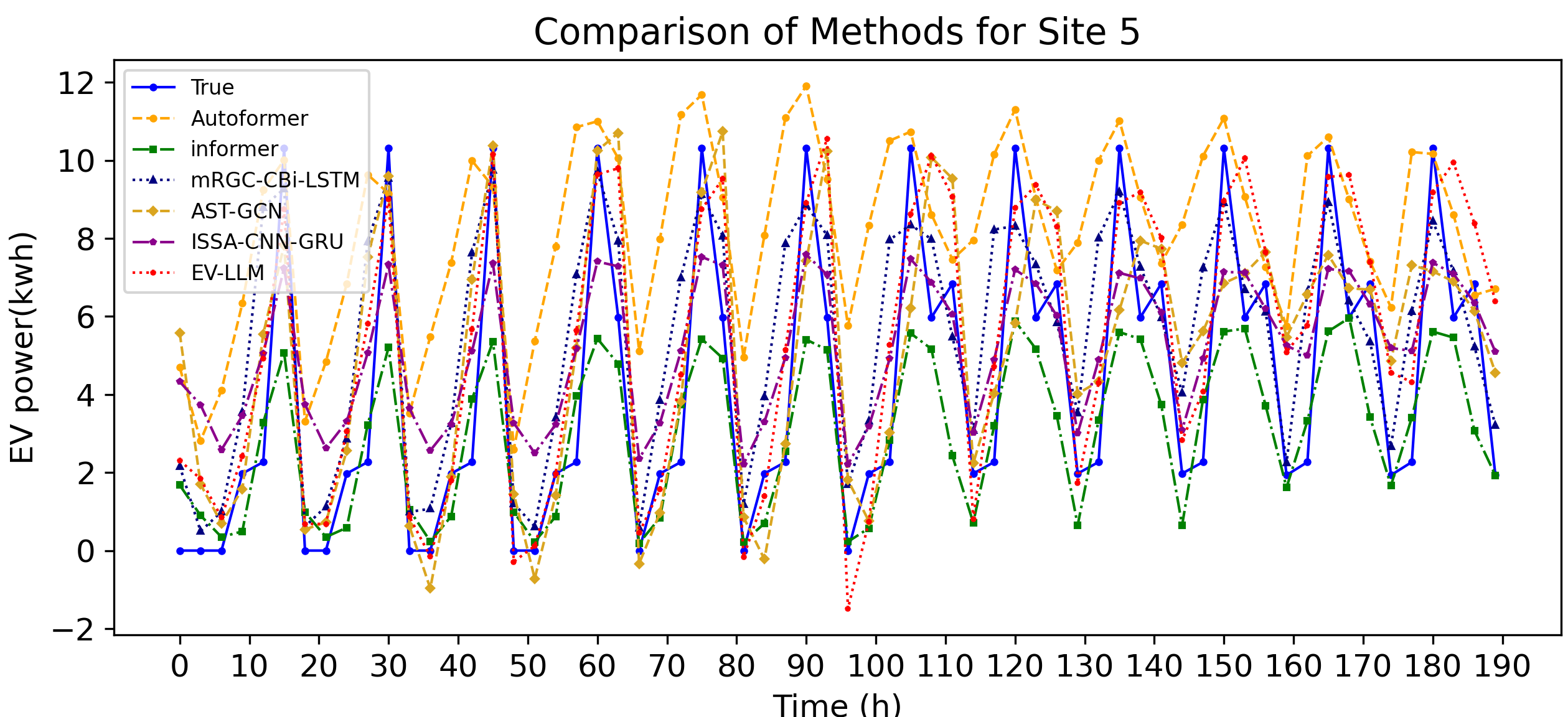}
                \caption{Prediction results of benchmark methods for the $16^{th}$ hour}
                \label{fig:step16}
        \end{figure}

To demonstrate the performance of each forecasting method, we visualize the prediction results for a specific charging station. We randomly select Station 5 and plot the prediction results for the 
and $16^{th}$ hours in 
Fig. \ref{fig:step16}.





As shown in Fig. \ref{fig:step16}, the prediction curve generated by our proposed EV-LLM model aligns more closely with the actual trend compared to the other methods. It accurately captures peak charging times and values, which is crucial for the effective operation of charging stations. These results reflect that our proposed EV-LLM framework enables the LLMs to adapt effectively to the specific task of EV charging load forecasting, and the integration of multimodal information significantly enhances prediction accuracy.

\subsection{Ablation Study}

To further validate the proposed model, we conduct ablation experiments from various perspectives, summarized in Table \ref{tab:abl}. The baseline model, EV-LLM, is derived from preliminary experiments. In contrast, EV-LLM (GPT2) replaces LLAMA2-7b with GPT2, while EV-LLM (w/o prompt) removes the prompt, and EV-LLM (w/o GCN) excludes the GCN module. Additionally, EV-LLM (mr 10\%/20\%) represents test results with 10\%/20\% of input data missing, which were handled through interpolation.

\begin{table}[ht]
  \centering
  \caption{Results of Ablation Studies}
  \label{tab:abl}
  \begin{tabular}{c|c|c} 
    \hline 
    Model & Average MAE& Average RMSE  \\ \hline 
    EV-LLM& 1.7288& 2.8960\\ 
    EV-LLM(GPT2) & 1.8487 & 3.0178\\
    EV-LLM(w/o prompt) & 1.9049 & 3.0425 \\ 
    EV-LLM(w/o GCN) & 1.9507 & 3.0721\\
    EV-LLM(mr 10\%) & 1.7478 & 2.9090\\ 
    EV-LLM(mr 20\%) & 1.7758 & 2.9426 \\  \hline
  \end{tabular}
\end{table}

The results indicate that replacing LLAMA2 with GPT2 leads to a reduction in prediction accuracy, suggesting that a larger LLM enhances performance. Furthermore, the inclusion of both the prompt and GCN modules significantly boosts accuracy, as their removal results in a decline in performance. Notably, EV-LLM demonstrates strong robustness to missing data, with moderate data loss having minimal impact on prediction performance.

\section{Conclusion}
In this paper, we propose a novel forecasting method, EV-LLM, for the spatiotemporal prediction of EV charging loads. EV-LLM effectively combines the strengths of GCNs in spatiotemporal feature extraction with the generalization capabilities of LLMs. This integration enables robust data mining and feature extraction across multimodal and multidimensional datasets, allowing for the incorporation of historical charging data, weather information, and relevant textual descriptions. Validation using charging data from 10 stations in California demonstrates that EV-LLM significantly outperforms other benchmark methods in forecasting accuracy, achieving lower MAE and RMSE.

Furthermore, future work will focus on applying EV-LLM to broader regions and diverse scenarios. Additionally, we aim to further optimize the model to accommodate larger datasets and address more complex environmental factors.



\bibliographystyle{IEEEtran}
\bibliography{ref}

\begin{thebibliography}{10}
\providecommand{\url}[1]{#1}
\csname url@samestyle\endcsname
\providecommand{\newblock}{\relax}
\providecommand{\bibinfo}[2]{#2}
\providecommand{\BIBentrySTDinterwordspacing}{\spaceskip=0pt\relax}
\providecommand{\BIBentryALTinterwordstretchfactor}{4}
\providecommand{\BIBentryALTinterwordspacing}{\spaceskip=\fontdimen2\font plus
\BIBentryALTinterwordstretchfactor\fontdimen3\font minus
  \fontdimen4\font\relax}
\providecommand{\BIBforeignlanguage}[2]{{%
\expandafter\ifx\csname l@#1\endcsname\relax
\typeout{** WARNING: IEEEtran.bst: No hyphenation pattern has been}%
\typeout{** loaded for the language `#1'. Using the pattern for}%
\typeout{** the default language instead.}%
\else
\language=\csname l@#1\endcsname
\fi
#2}}
\providecommand{\BIBdecl}{\relax}
\BIBdecl

\bibitem{starke2023improving}
M.~Starke, M.~Chinthavali, N.~Kim, T.~Carroll, F.~Tuffner, B.~Varghese,
  C.~Rieger, K.~Rohde, and T.~Pennington, ``Improving resiliency for electric
  vehicle charging,'' in \emph{2023 IEEE Power \& Energy Society General
  Meeting (PESGM)}.\hskip 1em plus 0.5em minus 0.4em\relax IEEE, 2023, pp.
  1--5.

\bibitem{su2017ev}
S.~Su, X.~Lin, H.~Zhang, H.~Zhao, H.~Li, and Z.~Li, ``Spatial and temporal
  distribution model of electric vehicle charging demand,'' \emph{Proceedings
  of the CSEE}, vol.~37, no.~16, pp. 4618--4629, 2017.

\bibitem{multi_info}
M.~Zhang, Q.~Sun, and X.~Yang, ``Electric vehicle charging load prediction
  considering multi-source information real-time interaction and user regret
  psychology,'' \emph{Power System Technology}, vol.~46, no.~2, 2022.

\bibitem{yaofang}
S.~C. Fang~Yao, Junhao~Tang and X.~Dong, ``A method for electric vehicle
  charging load forecasting based on the issa-cnn-gru model,'' \emph{Power
  System Protection and Control}, vol.~51, no.~16, pp. 158--167, 2023.

\bibitem{yao2024ev}
Y.~Zhang, C.~Liu, X.~Rao, X.~Zhang, and Y.~Zhou, ``Electric vehicle charging
  load prediction based on dynamic adaptive graph neural network,''
  \emph{Autom. Electr. Power Syst}, vol.~48, pp. 86--93, 2024.

\bibitem{wang2023predicting}
S.~Wang, A.~Chen, P.~Wang, and C.~Zhuge, ``Predicting electric vehicle charging
  demand using a heterogeneous spatio-temporal graph convolutional network,''
  \emph{Transportation Research Part C: Emerging Technologies}, vol. 153, p.
  104205, 2023.

\bibitem{jin2023time}
M.~Jin, S.~Wang, L.~Ma, Z.~Chu, J.~Y. Zhang, X.~Shi, P.-Y. Chen, Y.~Liang,
  Y.-F. Li, S.~Pan \emph{et~al.}, ``Time-llm: Time series forecasting by
  reprogramming large language models,'' \emph{arXiv preprint
  arXiv:2310.01728}, 2023.

\bibitem{wu2024stellm}
T.~Wu and Q.~Ling, ``Stellm: Spatio-temporal enhanced pre-trained large
  language model for wind speed forecasting,'' \emph{Applied Energy}, vol. 375,
  p. 124034, 2024.

\bibitem{ilharco2022patching}
G.~Ilharco, M.~Wortsman, S.~Y. Gadre, S.~Song, H.~Hajishirzi, S.~Kornblith,
  A.~Farhadi, and L.~Schmidt, ``Patching open-vocabulary models by
  interpolating weights,'' \emph{Advances in Neural Information Processing
  Systems}, vol.~35, pp. 29\,262--29\,277, 2022.

\bibitem{asudani2023impact}
D.~S. Asudani, N.~K. Nagwani, and P.~Singh, ``Impact of word embedding models
  on text analytics in deep learning environment: a review,'' \emph{Artificial
  intelligence review}, vol.~56, no.~9, pp. 10\,345--10\,425, 2023.

\bibitem{kim2024spatial}
H.~J. Kim and M.~K. Kim, ``Spatial-temporal graph convolutional-based recurrent
  network for electric vehicle charging stations demand forecasting in energy
  market,'' \emph{IEEE Transactions on Smart Grid}, 2024.

\end{thebibliography}
\end{document}